\documentclass[aps,prl,twocolumn,superscriptaddress,nofootinbib,showpacs]{revtex4-1}
\pdfoutput=1 
\usepackage{amsmath,latexsym,amssymb,graphicx,hyperref,color,enumerate,url}
\usepackage{slashed}

\hypersetup{colorlinks,citecolor= nicegreen,linkcolor= nicered}
\usepackage{color}
\definecolor{nicered}{rgb}{0.7,0.1,0.1}
\definecolor{nicegreen}{rgb}{0.1,0.5,0.1}

\begin{document}
\addtolength{\belowdisplayskip}{-.2ex}       
\addtolength{\abovedisplayskip}{-.2ex}       

\title{Dark Matter as the Trigger of Strong Electroweak Phase Transition}

\author{Talal Ahmed Chowdhury}
\affiliation{SISSA, Trieste, Italy}
%
\author{Miha Nemev\v sek}
\affiliation{ICTP, Trieste, Italy}
\affiliation{J.\ Stefan Institute, Ljubljana, Slovenia}
\author{Goran Senjanovi\'c}
\affiliation{ICTP, Trieste, Italy}
%
\author{Yue Zhang}
\affiliation{ICTP, Trieste, Italy}

\date{\today}

\begin{abstract}
\noindent
In this Letter, we propose a new possible connection between dark matter relic density and baryon asymmetry of the universe. The portal between standard model sector and dark matter not only controls the relic density and detections of dark matter, but also allows the dark matter to trigger the first order electroweak phase transition. We discuss systematically possible scalar dark matter candidates, starting from a real singlet to arbitrary high representations. We show that the
simplest realization is provided by a doublet, and that strong first-order electroweak phase transition implies a lower bound on the dark matter direct detection rate. The mass of dark matter lies between 45 and 80 GeV, allowing for an appreciable invisible decay width of the Standard Model Higgs boson, which is constrained to be lighter than 130 GeV for the sake of the strong phase transition.
 
\end{abstract}

\maketitle

\noindent{\sf\bfseries Introduction.}
The existence of dark matter (DM) has now been well established in astrophysical and cosmological observations, and it is known to constitute around twenty percent of the total energy density in the universe. Many experiments have been setup underground or in the sky in order to probe DM interactions.

The most popular DM candidate is a stable weakly-interacting massive particle (WIMP)~\cite{Scherrer:1985zt}. The correct thermal relic density is obtained if its mass lies near the electroweak scale, thus possible connections to electroweak symmetry breaking could be conjectured. Another possibility is the asymmetric dark matter~\cite{Nussinov:1985xr} whose origin can be related to the baryogenesis processes in the universe. This scenario has been revived recently in different incarnations~\cite{Kaplan:2009ag}, due to the observation that DM and baryon relic densities are of the same order of magnitude, and also partly inspired by the hints of a light GeV scale DM from direct detection experiments. One of the common features of the above listed work is, they all resort to baryon/lepton number violations beyond the SM. However, baryon number is known not to be exact in the SM~\cite{'tHooft:1976up} and it is possible to have baryon number violating process happening efficiently at a high temperature~\cite{Klinkhamer:1984di, Kuzmin:1985mm} if the minimal Higgs sector is extended.

In this Letter, we propose a scenario, where the portal between the standard model (SM) and dark sectors not only gives correct relic density and facilitates direct/indirect detections of DM, but also allows the DM to play an important role in the electroweak baryogenesis. Based on this picture, a new connection could be built between symmetric dark matter and asymmetric baryonic matter.



An essential ingredient for successful electroweak baryogenesis is
the existence of a strong enough first order phase transition. The finite temperature Higgs potential should contain a term proportional to $\phi^3 T$. Such a term exists in the SM but is not large enough.
If the DM were to do the job, it would have to be a scalar particle. Namely, fermions never contribute to the cubic term at one loop level. The question then is which scalar representation to take. 
As shown below, representations with integer weak isospin cannot work. 
In contrast, the half-integer representations are perfectly capable of playing this role.
We will opt for the most natural possibility of isospin one-half and leave a systematic study of higher multiplets for a future study.

We focus on a realization of such picture where the DM is identified as an inert doublet scalar~\cite{Barbieri:2006dq}. This possibility has been extensively pursued in recent years, and moreover, it follows naturally~\cite{Melfo:2011ie, Martinez:2011ua} in the case of mirror families~\cite{GellMann:1976pg, Wilczek:1981iz, Senjanovic:1984rw, Bagger:1984rk}, the old
dream of parity restoration of Lee and Yang~\cite{Lee:1956qn}. On the other hand, the extra Higgs doublet has also been amply used as a simple way of achieving first-order electroweak phase transition~\cite{Turok:1990in}.

We show here that both of the above can be achieved simultaneously.
This implies a lower bound on the DM direct detection rate, an exciting result in view of the upper limit from the Xenon experiment. Moreover, the masses of the charged and the pseudo-scalar component of the inert scalar doublet are constrained to be almost the same and lie in a rough window 270$-$350\,GeV, while the DM mass is small,
between 45$-$80\,GeV.
The SM Higgs boson mass is constrained to be lighter than 130\,GeV, which still allows for a large invisible decay width.
These are the main results of our paper.

There is also the need of having enough CP violation for the baryon asymmetry, which is an independent question and will be commented on in the outlook. In any case, the study of the phase transition is an important issue in itself.

\noindent{\sf\bfseries The inert doublet model.}
The existence of another scalar doublet is one of the most natural possibilities in the SM. Once introduced, it is natural to ask it to be the DM candidate~\cite{Barbieri:2006dq}. The price is quite high, for many couplings, including the new Yukawas, must be exceedingly tiny in order to guarantee sufficient stability of the DM. It is technically natural though, due to a discrete $Z_2$ symmetry that protects these couplings to remain small. One should keep in mind that the symmetry need not be exact since decaying DM is a viable possibility. For the sake of simplicity, we ignore the possible tiny breaking of $Z_2$, since it does not affect our discussion.

With the exact $Z_2$ symmetry $\Phi\to\Phi$, $D\to-D$, the Higgs potential takes a simple form
%
\begin{equation}
\label{eqPotential}
\begin{split}
V_0 = - \mu_\Phi^2 |\Phi|^2 + \mu_D^2 |D|^2 + \lambda_\Phi |\Phi|^4 + \lambda_D |D|^4  \\
 + \lambda_3 |\Phi|^2 |D|^2 + \lambda_4 |\Phi^\dagger D|^2 + \frac{\lambda_5}{2} \left[ (\Phi^\dagger D)^2 + \text{h.c.} \right] , 
\end{split}
\end{equation}
where $\Phi$ stands for the usual Higgs doublet and $D$ is the inert doublet, whose lightest component (assumed neutral) is to be identified as the DM.

Throughout the paper, we focus on the parameter space where $D$ does not acquire a vacuum expectation value (vev). The electroweak gauge symmetry is broken by the usual Higgs doublet $\Phi$ only~\cite{Hambye:2007vf}, which preserves the above $Z_2$ symmetry. At zero temperature, the mass spectra of components in the inert doublet are
\begin{equation}
\label{mass}
\begin{split}
m_S^2 = \mu_D^2\!+\!\frac{\lambda_S}{2}   v^2, \ 
m_A^2 = \mu_D^2\!+\!\frac{\lambda_A}{2}  v^2,  \ 
m_C^2 = \mu_D^2\!+\!\frac{\lambda_C}{2}  v^2. 
\end{split}
\end{equation}
Here, we define $\lambda_S=\lambda_3 + \lambda_4 + \lambda_5$, $\lambda_A =\lambda_3 + \lambda_4 - \lambda_5$
and $\lambda_C=\lambda_3$. We choose $S$ to be the DM candidate, which implies $\lambda_5<0$, 
$\lambda_4+\lambda_5<0$.

\smallskip
\noindent{\sc \small Electroweak Precision Tests.} 
Since the scalar contribution to the $S$ parameter is small, the crucial rule is played by the $T$ parameter, which gets a correction due to the inert doublet~\cite{Barbieri:2006dq}
\begin{eqnarray}
\Delta T \approx \frac{1}{24\pi^2 \alpha v^2} (m_C - m_A)(m_C-m_S) \ .
\end{eqnarray}
The SM itself, as is well known, prefers a light Higgs boson. In~\cite{Barbieri:2006dq}, the Higgs was taken to be heavy in order to improve naturalness, which then required substantial $\Delta T$ or a non-degenerate spectrum of the inert doublet states.

On the other hand, we are motivated by the first-order phase transition, which favors as light SM Higgs as possible, which in turn implies $\Delta T \approx 0$. There are two possibilities to achieve this: a) $m_S< m_C \approx m_A$ or b) $m_S\approx m_C<m_A$. It will turn out that $S$, as the DM, has to be lighter than about 80\,GeV. It is not completely clear whether the possibility b) is in accord with the experiment~\cite{Pierce:2007ut}. For this reason, in the bulk of this letter we pursue possibility a), where $A$ and $C$ are quite degenerate~\cite{Dolle:2009fn} or equivalently $\lambda_4 \approx \lambda_5$.

\smallskip
\noindent{\sc \small Scalar Doublet as Dark Matter.}
We consider $S$ as a thermal DM candidate, as in the conventional WIMP picture. The main processes governing the freeze out include annihilation into gauge bosons $W$, $Z$ and to fermions via the Higgs boson exchange. It has been shown in~\cite{Dolle:2009fn} that the WMAP already puts strong constraints on the viable parameter space.
The spin-independent direct detection cross section is~\cite{Shifman:1978zn, Barbieri:2006dq}
\begin{equation}
	\sigma_{\rm SI} = \frac{\lambda_S^2 f^2}{4\pi} \frac{\mu^2 m_N^2}{m_h^4 m_S^2} \ ,
\end{equation}
where $\mu = m_S m_N/(m_S+m_N)$.
The effective matrix element for the Higgs interaction with the nucleon is 
$f=f_{T_u}^{(N)}+f_{T_d}^{(N)}+f_{T_s}^{(N)}+ {2}/{9}$ and $m_N f_{T_q}^{(N)} \equiv \langle N|m_q \bar q q|N\rangle$ is the nucleon sigma term for light flavors. In order to be conservative, we have used the lowest values from a recent lattice calculation~\cite{Giedt:2009mr}.
We assume the other components $A$ and $C$ are much heavier than $S$, which as we show below, is the case when the phase transition constraint is taken into account. In this case, their roles in thermal freeze out and direct detection are safely negligible.

\begin{figure}[t!]
\centerline{\includegraphics[width=8.8cm]{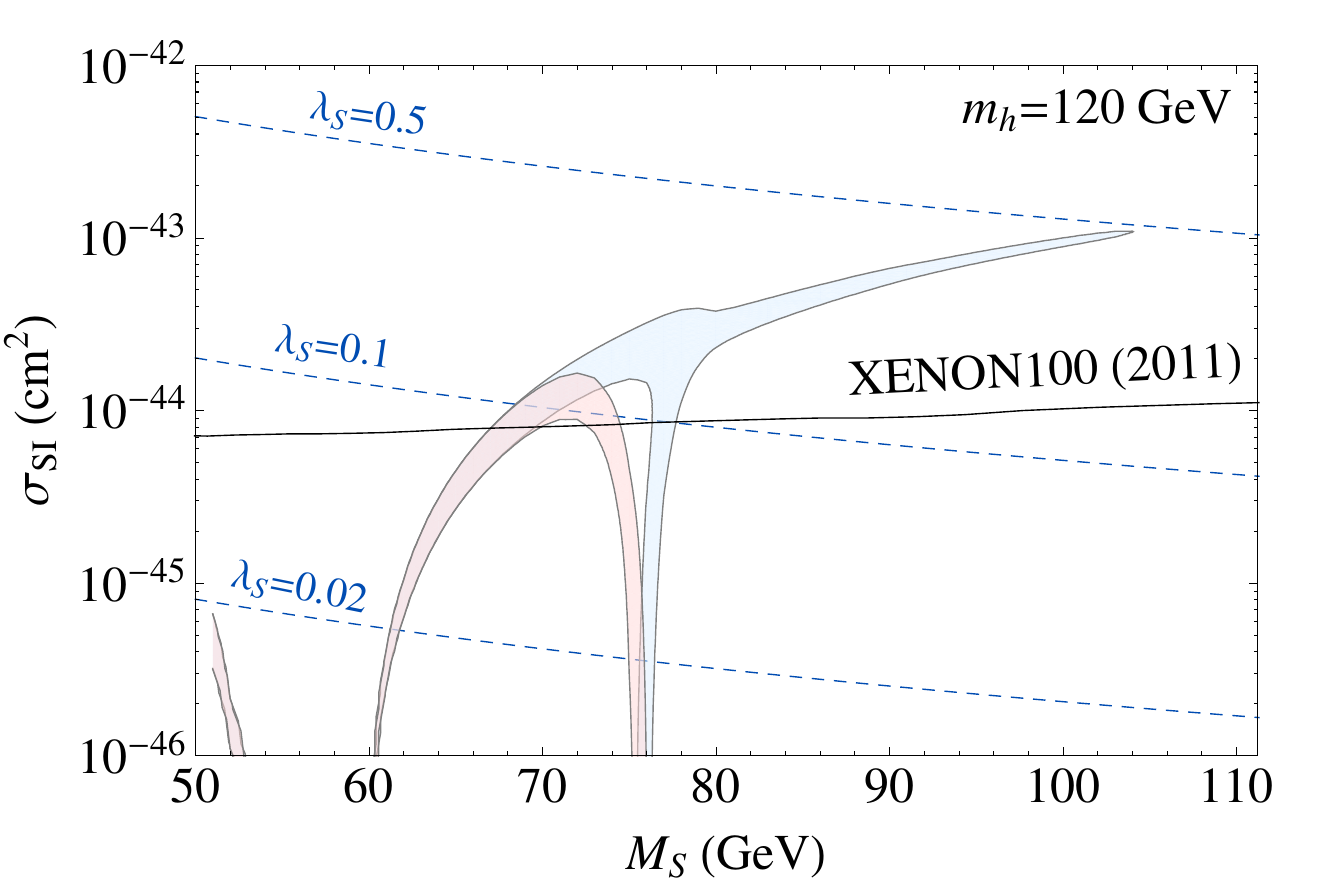}}
\caption{Spin-independent direct detection cross section on nucleon plotted as a function of the DM mass. Colored regions represent DM relic density favored by WMAP, $\Omega_{\rm DM}h^2 \in (0.085, 0.139)$ at $3\sigma$, for positive (red) and negative (blue) $\lambda_S$. We have taken SM Higgs mass $m_h=120\,$GeV. The lower limit on the direct detection cross section from Xenon100 experiment is shown by the black solid line. Also shown in the figure are the dashed curves for constant $|\lambda_S|$.}\label{relic}
\end{figure}

The combined limit from both relic density and direct detection experiments demands the DM mass to lie approximately
between 45$-$80\,GeV~\cite{Melfo:2011ie}. This result is shown in Fig.~\ref{relic} for the SM Higgs mass equal to 120\,GeV.
In the above favored DM mass window, Xenon\,100 has also put tight constraint on the interaction between DM and 
the Higgs boson,
\begin{equation}
 \label{lambdas}
 \lambda_S \lesssim 0.1 \ .
\end{equation}
A useful constraint follows from the above results. The mass spectrum in Eq.(\ref{mass}) leads to $\lambda_3\approx -2\lambda_4 \approx -2\lambda_5$, when $m_A\approx m_C\gg m_S$, up to corrections of order 10\,\%. Demanding the SM Higgs boson to remain light up to a few hundred GeV energy scale, i.e., $\Delta \lambda_\Phi \sim (3\lambda_3/2) \log (\mu/m_h) \lesssim \lambda_\Phi$ sets a relevant upper bound on the couplings between the two scalar doublets. In particular, we find $\lambda_C \equiv \lambda_3\lesssim 4$. Therefore,
\begin{eqnarray}
m_A\approx m_C\lesssim 350\, {\rm GeV} \ .
\end{eqnarray}
The mono-chromatic gamma ray line from DM annihilation in the galaxy could also serve as a promising indirect detection of DM. In this model, the flux is predicted~\cite{Melfo:2011ie} to lie only a factor of 4$-$5 below the current Fermi-LAT limit.

\smallskip
\noindent{\sc \small First-order Phase Transition.}
At high temperature, the Higgs potential improved after the so called daisy re-summation takes the following form~\cite{Turok:1991uc, Cline:1996mga}
\begin{eqnarray}
	V_{\text{tot}}\!&\approx&\!\frac{1}{4}\lambda_\Phi \phi^4\!+\!\frac{1}{2} \left[ -\mu_\Phi^2\!+\!a\, \frac{T^2}{12} \right] \phi^2\!-\!\frac{T}{12\pi} \sum_{\rm B} n_B\, m_B^3(\phi, T) \nonumber \\
	&& \ \  - \sum_B n_B \frac{m_B^4(\phi, T)}{64\pi^2} \left[ \log \frac{m_B^2(\phi, T)}{T^2} - 5.40762 \right]  \\
	&& \ \ +  \sum_F n_F \frac{m_F^4(\phi, T)}{64\pi^2} \left[ \log \frac{m_F^2(\phi, T)}{T^2} - 2.63503\right],  \nonumber
\end{eqnarray} 
where $M_{W,Z} (\phi) = ({\phi}/{v}) M_{W,Z}$, $m_t (\phi) = ({\phi}/{v}) m_t$ and the numbers of degrees of freedom for particles relevant here $\{W^\pm, Z^0, h, G^0, G^\pm, S, A, C^\pm, t\}$ are $\{4, 2, 1, 1, 2, 1, 1, 2, 12\}$. 
The vev of the Higgs field $\phi=v=246\,$GeV at zero temperature.
The scalar thermal masses are~\cite{Cline:1995dg, Mohapatra:1979vr}
\begin{eqnarray}\label{thermalmass}
&&m_h^2(\phi, T) = m_{G^\pm, G^0}^2(\phi, T)  \approx - \mu_\Phi^2 + a \, \frac{T^2}{12}  +  3 \lambda_\Phi \phi^2 \ , \nonumber \\
&&m_i^2(\phi, T) \approx \left(\mu_D^2 + b \, \frac{T^2}{12} \right) + \frac{1}{2} \lambda_i \phi^2 \ , (i=S, A, C),
\end{eqnarray}
with $a = 6\lambda_\Phi +  2\lambda_3 + \lambda_4 + {(9g^2 + 3g'^2)}/{4} + 3 y_t^2$ and  
$b = 6 \lambda_D + 2\lambda_3+\lambda_4 + {(9g^2 + 3g'^2)}/{4}$.

The criterion for having a strong first-order electroweak phase transition is $v_c/T_c\gtrsim1$ at the critical point, which calls for a large cubic thermal potential $\phi^3 T$. In the inert doublet model the new scalars are expected to play this crucial role with a sufficiently large coupling to the SM Higgs boson, $\lambda_i\sim\mathcal{O}(1)$. 
In the previous section, we showed that direct detection implies $\lambda_S\lesssim 0.1$. Therefore, only large $\lambda_A$ and $\lambda_C$ could do the job, which implies that the corresponding pseudo scalar and charged components are heavy at zero temperature.

\begin{figure}[t!]
\centerline{\includegraphics[width=8.8cm]{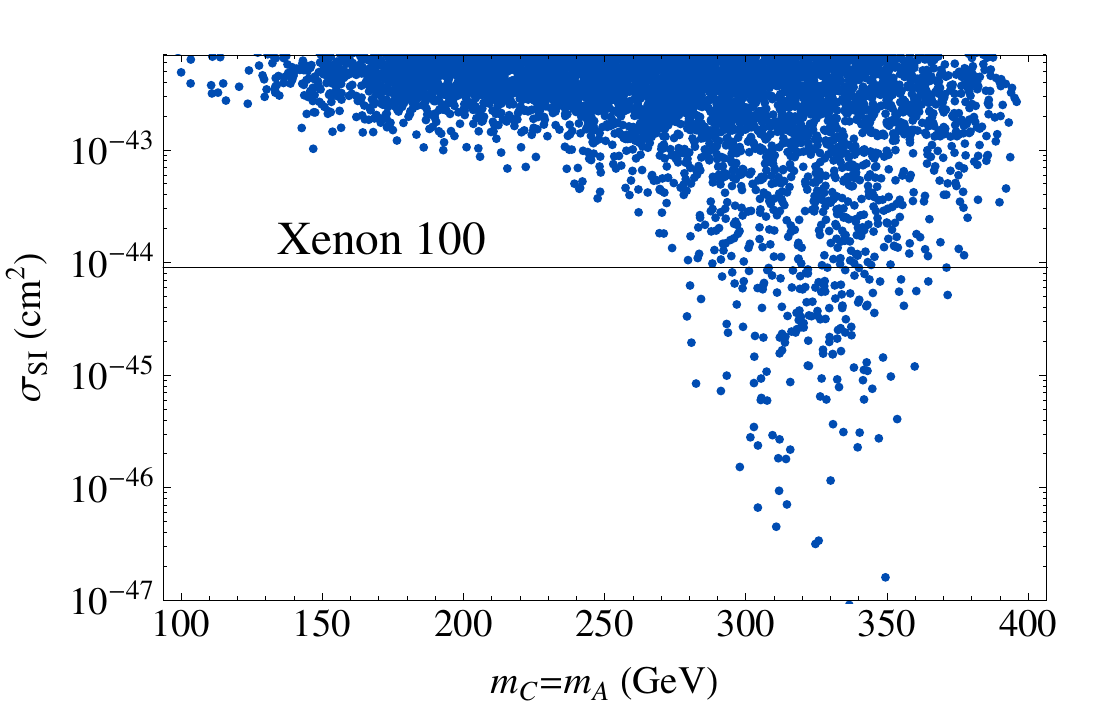}}
\caption{Correlation between spin-independent direct detection cross section and  the mass of the charged scalar, after imposing the strong electroweak transition condition $v_c/T_c>1$. We have scanned the parameter space: $m_h\in(115, 200)\,$GeV, $m_S\in(40, 80)\,$GeV, $m_A\in(100, 500)\,$GeV, $m_C\in(m_A-10\,{\rm GeV}, m_A+10\,{\rm GeV})$, $\lambda_S\in(0, 1)$ and $\lambda_D\in(0,3)$. We veto points where the thermal mass of $A$ or $C$ exceeds $1.8\, T_c$, which would invalidate the high temperature expansion. }\label{xenon}
\end{figure}

The second crucial point to note is that the $\phi$-independent term in the thermal masses of $A$ and $C$ should be small enough in order not to dilute their contribution to the cubic term. Namely, the optimal situation is realized when the terms in the bracket of Eq.~(\ref{thermalmass})
\begin{equation}
\!\!\mu_D^2\!+\!\frac{T_c^2}{12}\!\left[ 6 \lambda_D\!+\!\frac{m_S^2\!+\!m_A^2\!+\!2m_C^2\!-\!4\mu_D^2}{v_c^2}\!+\!\frac{9g^2 + 3g'^2}{4} \right]\!,
\end{equation}
are minimized at the critical temperature. For a given mass spectrum this means that there is a window for $\mu_D^2$ where the phase transition could be strongly first order. Since $S$ shares the same $\mu_D^2$ contribution to the mass as its heavier partners, i.e., $\mu_D^2 = m_S^2 - \lambda_S v^2/2$, it in turn predicts a lower bound on the DM direct detection rate, as shown in Fig.~\ref{xenon}. 
This is an important result in view of the upper bound set by Xenon\,100, which constrains the masses of $A$ and $C$ to lie in a window between 270$-$350\,GeV.

This mass window of heavy scalar masses can be probed by the LHC. In particular, the pseudo-scalar component $A$ can be produced in association with the DM $S$, and leads to dilepton final state with missing energy. Here, the preferred mass difference between $A$ and $S$ is larger than the sample values studied in~\cite{Dolle:2009ft}, which makes it easier to be distinguished from the SM background by imposing a harder cut on the missing energy.

\begin{figure}[t!]
\centerline{\includegraphics[width=4.6cm]{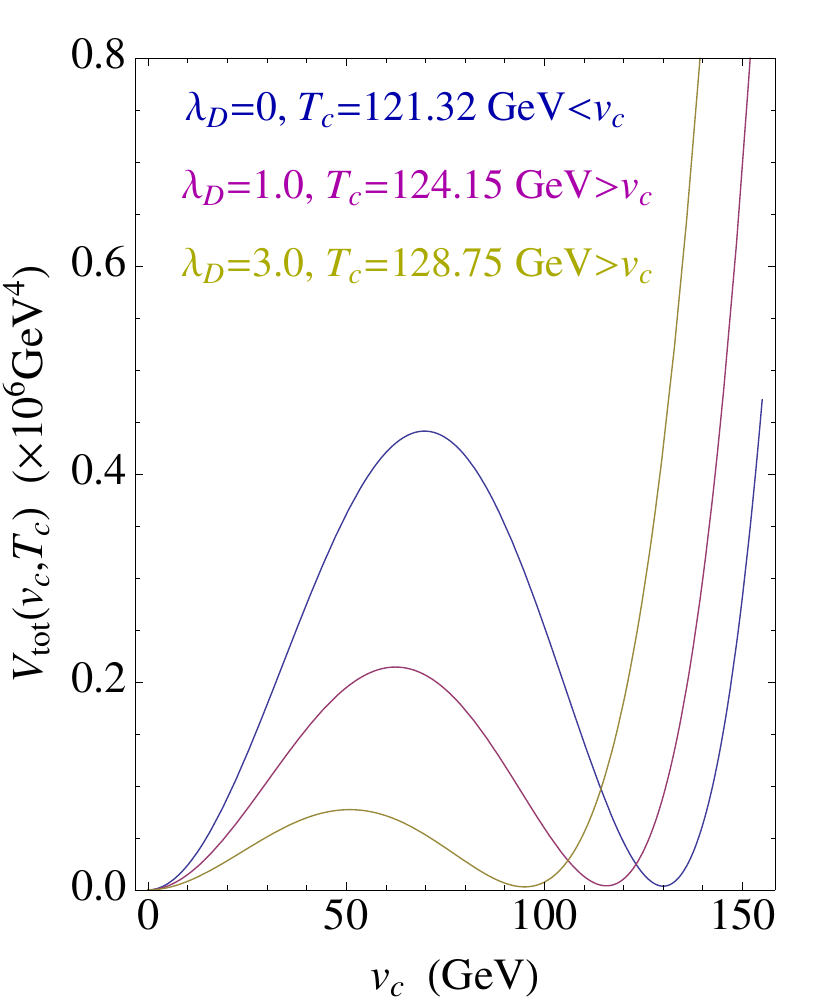}\hspace{-0.6cm}
\includegraphics[width=4.504cm]{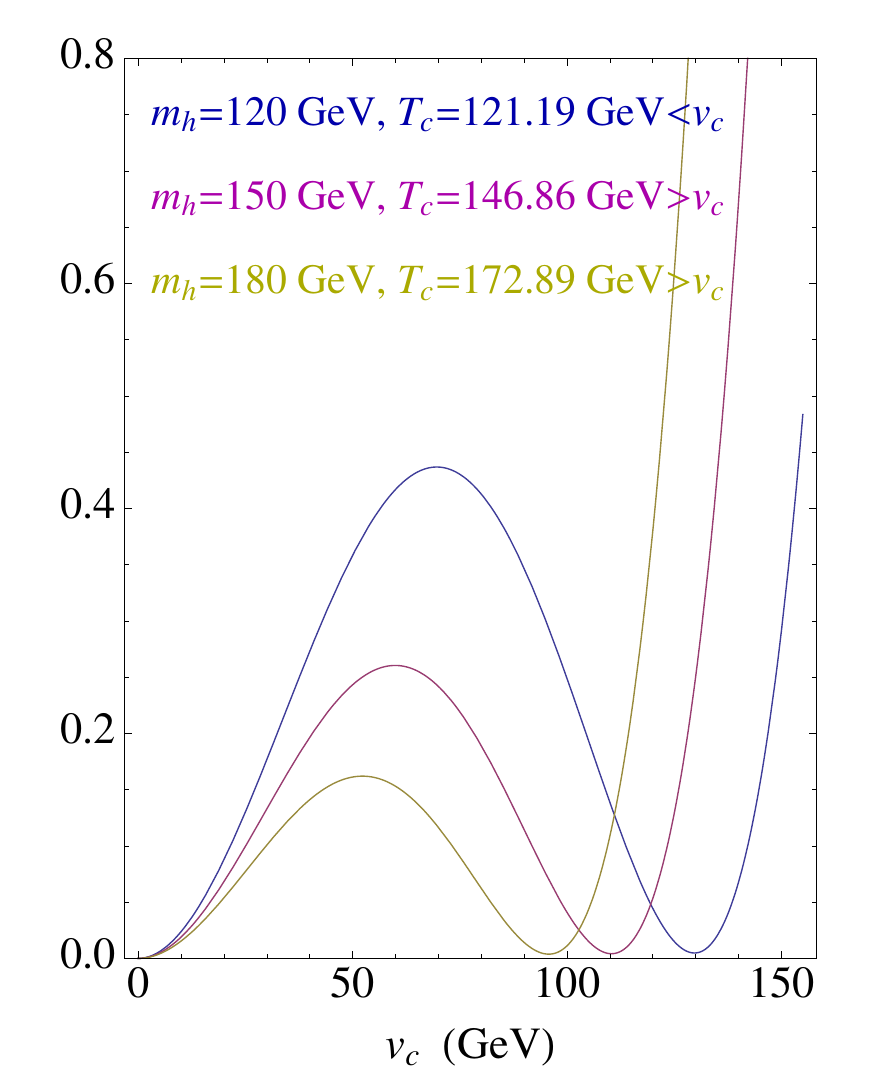}}
\caption{Shape of the Higgs potential at the critical temperature and its dependence on different choices of parameters: DM self-interaction $\lambda_D$ (left panel) and SM Higgs boson mass $m_h$ (right panel). While varying $\lambda_D$, we have fixed $m_h=120\,$GeV, $m_S=60\,$GeV, $m_A=m_C=300\,$GeV and while varying $m_h$, we have fixed $\lambda_D=0$, $m_S=76\,$GeV, $m_A=m_C=300\,$GeV, respectively.}\label{Tc}
\end{figure}

We also show in Fig.~\ref{Tc} the dependence of critical temperature and the Higgs vev on the DM self coupling 
$\lambda_D$ and the SM Higgs mass. We find that by increasing the DM self-interaction or enlarging the SM
Higgs mass both reduce $v_c$ and increase $T_c$, and thus weaken the strength of the phase transition.
In particular, $T_c$ increases very quickly with the Higgs mass and we find an upper bound on the Higgs boson
mass~\footnote{In this model the mass of dark matter lies in a window that could open appreciable invisible decay channel of the Higgs, controlled by the DM-Higgs coupling $\lambda_S$. Therefore, the Higgs branching ratios can deviate from SM prediction.}
\begin{equation}
m_h\lesssim 130\,{\rm GeV} \ .
\end{equation}
In short, the scalar doublet DM can trigger the strong electroweak phase transition, as long as it is light, below 80 GeV or so, and its partners end up being heavier.

\smallskip
\noindent{\sc \small What about Uncertainties?}
So far, we have worked in the improved one-loop approximation for the effective potential at non-zero temperature, and so one can question its reliability at higher orders in perturbation theory. In the examples studied up to now, such as MSSM, it turns out the two-loop effects~\cite{Espinosa:1996qw} only help to strengthen the phase transition. Similarly, the non-perturbative lattice simulations tend to do the same over the perturbative results~\cite{Laine:2000xu}.

Another uncertainty lies in the possibly effect of the magnetic field during the phase transition~\cite{Klinkhamer:1984di}. The size of the magnetic field has been up to now only roughly estimated~\cite{Baym:1995fk}, thus its effect is not completely clear. It was argued recently~\cite{DeSimone:2011ek} though, in the context of the MSSM, that it may have an impact on the upper limit of the Higgs mass. 

Recently, the issue of gauge invariance has been brought up~\cite{Patel:2011th}. It is claimed that one may again need a complete two-loop finite-temperature effective potential for this purpose.

\smallskip
\noindent{\sf\bfseries Why not a singlet?} 
Before turning to higher representations, let us discuss explicitly the case of the singlet DM. After all, this is 
a simpler possibility with fewer couplings and thus more constrained.
In fact, it fails to do the job. More precisely, while the singlet by itself can actually help the phase transition to be of the first order~\cite{Anderson:1991zb}, it cannot simultaneously be the DM~\cite{Espinosa:2008kw}, and vice versa.

What happens is the following. In this case, there is only one coupling with the Higgs and $\lambda_A \equiv \lambda_C \equiv \lambda_S$.  
We survey all the points in Fig.~\ref{xenon} and find they all satisfy $\lambda_{A,C}\gtrsim1$. On the other hand,
direct detection, as shown in Eq.~(\ref{lambdas}), constrains this coupling to be much smaller than what is
needed to trigger a strong first-order phase transition. The failure of the real singlet thus makes the choice of the inert
doublet scalar the simplest one.

One can further extend the real scalar singlet case to a complex one. It was shown~\cite{Barger:2008jx} that the
double job of dark matter and strong electroweak phase transition can be achieved in this case.

On the other hand, the scalar singlet could be the carrier of the force between the SM sector and the
dark matter one~\cite{Das:2009ue}, instead of being DM itself. Such a singlet can actually trigger~\cite{Das:2009ue}
the first order phase transition. This can be successfully embedded~\cite{Carena:2011jy} in the NMSSM.

\smallskip
\noindent{\sf\bfseries  Higher representation alternative?}
It could be appealing to resort to higher $SU(2)_L$ representations for DM candidate, 
since then there are fewer $Z_2$ odd couplings which destabilize them.

Let us start with integer isospin representations $\Delta$. In order to have a neutral particle, 
needed for the DM, they must have even hypercharge. Therefore, they only have
two gauge invariant terms with the SM Higgs, out of which only one can split their masses 
\begin{equation}
 \left(\Delta^\dagger T^a  \Delta \right)\left(\Phi^\dagger \sigma^a \Phi \right) \ ,
\end{equation}
where $T^a$ are the appropriate generators of $\Delta$. In the case of the real multiplet
with $Y=0$, the spectrum is degenerate, while in the case of the complex one, the mass
splits are proportional to the electromagnetic charge once the Higgs gets the vev. 
 
The former case works only for a heavy DM, above TeV, due to strong
co-annihilating effects on the relic density~\cite{Cirelli:2005uq}. This makes it too heavy 
to have an impact on the phase transition. The latter case implies degenerate real and
imaginary components of the neutral particle, which couple to the $Z$. Direct detection
limits can be evaded again with a TeV scale DM. In short, as remarked in the Introduction,
the integer isospin candidates fail to render the phase transition be first order.
 
How about higher half-integer isospin multiplets? A natural choice $Y=1/2$, accommodates
another term in the potential
\begin{equation}
  \left( \Delta^T T^a  \Delta \right) \left( \Phi^T \sigma^a \Phi \right)^* \ ,
\end{equation}
where we ignore for simplicity the $SU(2)$ conjugation. In general, this term splits the real and
imaginary neutral components and in principle allows for light DM and heavy enough
other states, just as in the case of the doublet discussed above. We will return to this
intriguing possibility in a future publication~\cite{future}.

\smallskip
\noindent{\sf\bfseries Outlook: what about genesis?}
Before closing let us comment on a few related issues.


\smallskip
\noindent{\sc \small Sources of CP Violation.}
Successful baryogenesis requires CP violation, not only the first order phase transition.  
It is easy to imagine new sources of CP violation, but the problem then arises as to whether the new
physics behind it affects the nature of the phase transition. In this sense, new fermions are more
welcome, at least in the perturbative regime, while adding
new scalars is less desirable since they may upset the first order phase transition. Of course, there is 
always a large parameter space where they are innocuous, since even tiny couplings with the SM particles
may give enough CP violation.

\smallskip   
\noindent{\sc \small Testing the Thermal History.}  As we showed, the first order phase transition implies a window
for particle masses, that can serve as a probe at colliders. Still, one would like to have a more direct consequence that reflects the cosmological dynamics. One intriguing possibility are the remnant gravitational waves~\cite{Kamionkowski:1993fg}.

\smallskip
\noindent{\sc  \small Pre-existing Baryon Asymmetry?}
In order to elucidate this possibility, let us recall the problem in the conventional picture:
the sphaleron erasure~\cite{Kuzmin:1985mm} of any original $B+L$ asymmetry, based on the conventional
assumption of symmetry restoration at high temperature.
  This, however, is not necessarily true in the BSM physics with enlarged Higgs sector: gauge symmetry 
may get even more broken at high temperature~\cite{Weinberg:1974hy, Mohapatra:1979vr}. For this to happen,
it suffices that the cross-term couplings between different scalar multiplets be negative. It may be worthwhile
to investigate this possibility; however it requires improving the perturbative results for next-to-leading effects
tend to be large~\cite{Bimonte:1996cw}. 
 
\medskip
\noindent{\sf\bfseries Acknowledgments.}
We are grateful to Alejandra Melfo for very useful discussions and suggestions. 
We thank James Cline, Andrea De Simone, Jose R. Espinosa, Mikko Laine, Michael Ramsey-Musolf and Francesco Riva for useful correspondence regarding the quantitative studies of strong electroweak phase transition. 
We acknowledge the BIAS institute for hospitality.
The work of G.S. is supported in part by the EU grant UNILHC-Grant Agreement PITN-GA-2009-237920. 
The work of YZ is supported in part by the National Science Foundation under Grant No. 1066293 and the hospitality of the Aspen Center for Physics.

\end{document}